\documentclass[conference]{IEEEtran}
\IEEEoverridecommandlockouts
\usepackage{cite}
\usepackage{amsmath,amssymb,amsfonts,bm}
\usepackage{algorithm}
\usepackage{algorithmic}
\usepackage{epsfig}
\usepackage{graphicx}
\usepackage{subfigure}
\usepackage{epstopdf}
\usepackage{stfloats}
\usepackage{url}
\usepackage[percent]{overpic}
\usepackage{xcolor}
\usepackage{rotating}
\usepackage{textcomp}

\newcommand{\norm}[1]{\left\|#1\right\|}
\def\BibTeX{{\rm B\kern-.05em{\sc i\kern-.025em b}\kern-.08em
    T\kern-.1667em\lower.7ex\hbox{E}\kern-.125emX}}

\begin{document}

\title{Radar Interferometry using Two Images with Different Resolutions}
\author{\IEEEauthorblockN{Huizhang~Yang,
	    Chengzhi~Chen,
        Shengyao~Chen,
        Feng~Xi and
        Zhong~Liu}
        \IEEEauthorblockA{Department of Electronic Engineering\\
        Nanjing University of Science \& Technology\\
Nanjing, Jiangsu, 210094, P. R. China}
        \thanks{\hrule \vspace{0.6em}This work was supported in part by the National Natural Science Foundation of China under Grants 61671245, 61571228.}}
\maketitle

\begin{abstract}
 Radar interferometry usually exploits two complex-valued radar images with the same resolution to extract terrain elevation information.
 This paper considers the interferometry using two radar images with different resolutions, which we refer to as dual-resolution radar interferometry.
 We find that it is feasible to recover a high-resolution interferogram from a high-resolution image and a low-resolution one.
 We formulate the dual-resolution interferometry into a compressive sensing problem, and  exploit the wavelet-domain sparsity of the interferogram to solve it.
 Due to the {speckle effect} in coherent radar imaging, the sensing matrix of our model is expected to have small mutual coherence, which guarantees  the performance of our method.
 In comparison with the conventional radar interferometry methods, the proposed method reduces the resolution requirement of radar image acquisition. It therefore can promote wide coverage, low sampling/data rate and storage cost.
 Numerical experiments on Sentinel-1 data are made to validate our method.
\end{abstract}


\IEEEpeerreviewmaketitle

\section{Introduction}

From its very beginning ideas \cite{Campbell1970Radar,Shapiro1972Lunar,Graham1974}, radar interferometry technique has been developed for more than four decades. This technique requires that the two radar images  have the same resolution after proper processing \cite{Bamler1999Synthetic}.  The resolution determines the fast- and slow-time bandwidths of the radar signal, and thus it also determines the rate of the sampling devices and the pulse repetition frequency (PRF).  Therefore, for high-resolution applications, the requirement of high sampling rate brings a big challenge to sampling devices, and it often needs to sacrifice coverage \cite{cumming} in spaceborne synthetic aperture radar (SAR). On the other hand, to obtain global digital elevation model (DEM), it needs to acquire large amounts of radar data, which puts huge burden on storage and down-link transmission. A example is the SRTM mission, by which the raw data collected is up to 9.8 Terabyte \cite{Bamler1999The}.

\begin{figure}
  \centering
  \includegraphics[width=9.0cm]{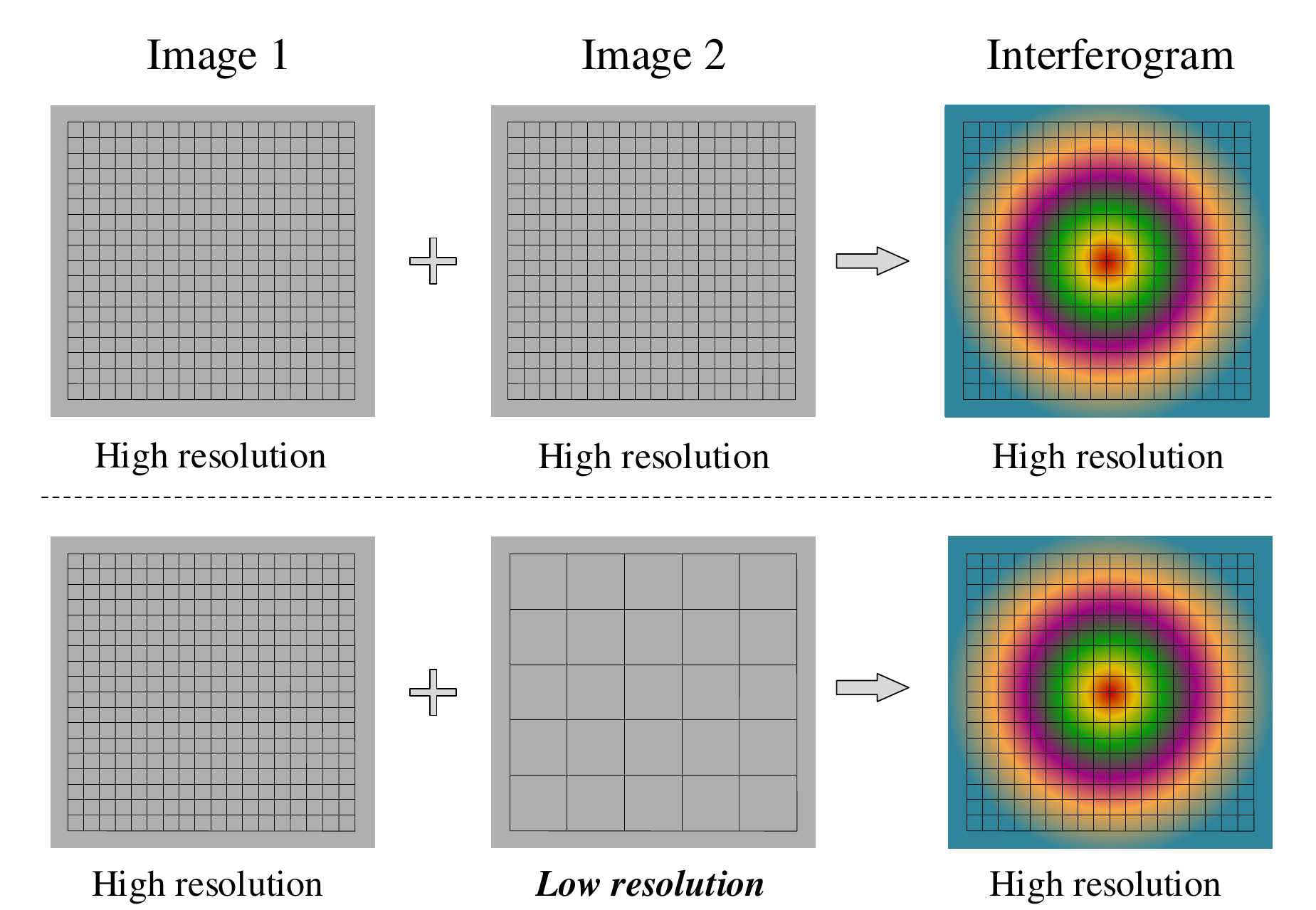}
  \caption{Conventional radar interferometry method (top) vs our method (bottom). The grids indicate the sizes of resolutions. Using our method, it is feasible to recover the high-resolution interferogram from a high-resolution image and a low-resolution one.}\label{insar_con}
\end{figure}

Motivated by these issues, this paper proposes a new radar interferometry method that involves two images with different resolutions: one image is high-resolution and the other is low-resolution. The proposed method, which we refer to as dual-resolution interferometry, can still extract high-resolution terrain elevation information, as illustrated in Fig. \ref{insar_con}. 
The low-resolution acquisition means low observation bandwidths in range and azimuth dimensions, and thus brings a reduction in sampling/data rate and storage cost, and promotes wide coverage or small revisit time. Specifically, for a single-pass mission such as SRTM, we can use a pair of high- and  low-resolution radars, and the low-resolution one can bring a reduction in sampling rate and storage cost.  For a repeat-pass mission such as  Sentinel-1, we can alternate the TOPS mode (low-resolution) and the stripmap mode (high-resolution) in two adjacent repeat cycles to reduce the revisit time.

Different from the conventional radar or SAR interferometry, we describe the interferometry with dual resolutions as a compressive sensing \cite{candes2008introduction} problem.
By exploiting the sparsity of interferogram in wavelet domain, the interferometric processing is transformed into a sparse recovery problem.
According to compressive sensing theory,  the sensing matrix of our model should have suitable properties, such as small mutual coherence \cite{Eldar2012Compressed}.
Thanks to the inherent randomness of the speckle effect in coherent radar imaging \cite{Frieden1976Laser,Goodman1976Some}, the sensing matrix of our model is random.
It  is expected to have small mutual coherence, and thus the  performance of our method will be guaranteed.
Numerical experiments on Sentinel-1 data shows that our method can still extract terrain elevation effectively even when the low resolution  is  only $1/5\times 1/5$ of the high resolution in range and azimuth dimensions.
Note that the proposed method can be incorporated into most existing SAR systems to simplify radar interferometry applications, so it may provide a better way for long-term Earth observation.

\section{Background}
Let us denote a complex-valued radar image by $z[n,l]$,  where $1\le n\le N$ and $1\le l\le L$ are integers indexing the image pixels.
The phase of  $z[n,l]$ can be decomposed into three components (ignore interference and noise) \cite{Ferretti2007InSAR}
\begin{equation}\label{}
	\phi[n,l]=\phi_{scat}[n,l]+\phi_{flat}[n,l]+\phi_{elev}[n,l],
\end{equation}
where $\phi_{scat}[n,l]$ is the scattering phase, $\phi_{flat}[n,l]$ and $\phi_{elev}[n,l]$ are the phases due to  flat Earth and   terrain elevation, respectively. To retrieve the  elevation information, radar interferometry exploits the phases of two radar images \cite{Bamler1999Synthetic}. Suppose that images  $z_1[n,l]$ and $z_2[n,l]$ are observed with the same frequency band but at slightly different look angles, then after coregistration,  they can be combined pixel by pixel
\begin{equation}\label{intf}
	z_{ifg}[n,l]=z^*_1[n,l]z_2[n,l].
\end{equation}
where $z^*_1[n,l]$ denote the conjugate of $z_1[n,l]$.
This operation results in a interferometric phase
 \begin{equation}\label{deltaphi}
  \phi_2[n,l]-\phi_1[n,l]= \Delta\phi_{flat}[n,l]+\Delta\phi_{elev}[n,l],
\end{equation}
under the assumption of fully correlation ($\phi_{scat}[n,l]$ is the same for the two images \cite{Bamler1999Synthetic}). Then by removing flat-Earth contribution using auxiliary orbit data, we have
	$\phi_{ifg}[n,l]=\Delta\phi_{elev}[n,l]$,
which contains the terrain elevation information.

For the pixel-by-pixel interferometry operation (\ref{intf}), it is required that the two images should have the same resolution after proper processing. Otherwise, each pixel pair will be decorrelated. Can we derive the terrain elevation from two images that have different resolutions? This is the problem we shall discuss in the reminder of this paper.

\section{ Dual-Resolution Model}
 In this section, we develop the mathematical model of the proposed dual-resolution radar interferometry problem.

\subsection{ Relations between Dual-resolution Images }

We consider this case: the first image $z_1[n,l]$ has the  bandwidth $B_{\tau}\times B_{\eta}$ in fast time $\tau$ and slow time $\eta$, and the second image $z^{\prime}_2[m,k]$ ($1\le m\le M$, $1\le k\le K$) has the  bandwidth $\alpha B_{\tau}\times \beta B_{\eta}$  with $\alpha,\beta\in(0,1)$. Then the first image has the (fast-time and slow-time) resolution $1/B_{\tau}\times 1/B_{\eta}$  while the second image has the reduced resolution $1/(\alpha B_{\tau})\times 1/(\beta B_{\eta})$. For simplicity, we assume that the two images are acquired using the stripmap mode with the same carrier frequency.

Based on the assumption of sinc point spread function $\mathrm{sinc}(B_{\tau}\tau)\mathrm{sinc}(B_{\eta}\eta)$ of bandlimited radar \cite{Bamler1999Synthetic,Li1990Studies}, the image $z^{\prime}_2[m,k]$ can be viewed as low-frequency part of the original image $z_2[n,l]$. Then  the spectrum of $z^{\prime}_2[m,k]$ and $z_2[n,l]$  can be aligned by common band filtering \cite{Ferretti2007InSAR} as
\begin{equation}\label{dftrelation}
	\mathcal{F} z^{\prime}_2[m,k]=\mathcal{S}_{\alpha,\beta}\mathcal{F} z_{2}[n,l],
\end{equation}
where the operator $\mathcal{F}$ is the two-dimensional discrete Fourier transform (2-D DFT) and $\mathcal{S}_{\alpha,\beta}$ is the 2-D lowpass common band filter.

Note that $\phi_2[n,l]= \phi_1[n,l]+\Delta\phi_{flat}[n,l]+\Delta\phi_{elev}[n,l]$ according to (\ref{deltaphi}). Then we can rewrite the image $z_2[n,l]$ as
\begin{equation}\label{ms}
	z_2[n,l]=\theta[n,l]u[n,l],
\end{equation}
where $\theta[n,l]=e^{j\phi_{1}[n,l]+j\Delta\phi_{flat}[n,l]}$ and $u[n,l]=|z_2[n,l]|e^{j\Delta\phi_{elev}[n,l]}$. Since the phase of $u[n,l]$ contains the interferometric information, we call it as interferogram in this paper.
Substituting (\ref{ms}) into (\ref{dftrelation}), we have
\begin{equation}\label{msdft}
	\mathcal{F} z^{\prime}_2[m,k]=\mathcal{S}_{\alpha,\beta}\mathcal{F} \left\{\theta[n,l]u[n,l]\right\}.
\end{equation}
 In practice, (\ref{ms}) and (\ref{msdft}) will be corrupted due to the noise introduced by atmosphere, temporal change and look angle. This mismodeling will be considered later.

\subsection{Underdetermined Model}
For ease of discussing interferogram formation, we shall express the linear model (\ref{msdft}) in matrix-vector form. To describe the involved DFT operations, we define four  DFT matrices, $\mathbf{F}_{M}$, $\mathbf{F}_{K}$, $\mathbf{F}_{N}$ and $\mathbf{F}_{L}$, where the subscripts denote the points of corresponding DFT operations. To describe the 2-D low-pass filter  $\mathcal{S}_{\alpha,\beta}$, let us define two submatrices $\mathbf{\Omega}_{\alpha}$ and $\mathbf{\Omega}_{\beta}$ of $N\times N$ and $L\times L$ identity matrices,  which extract the $\alpha \times \beta$ (in percentage) lowpass frequency in range and azimuth dimensions, respectively. Next, we  denote ${\mathbf{Z}^{\prime}_2}$, $\mathbf{\Theta}$ and $\mathbf{U}$ as the matrix forms of $z^{\prime}_2[m,k]$, $\theta[n,l]$ and $u[n,l]$, respectively. Then we can rewrite (\ref{msdft}) in matrix form
\begin{equation}\label{s3e2}
	{\mathbf{Z}^{\prime}_2}=\mathbf{F}_{M}^*\mathbf{\Omega}_{\alpha}\mathbf{F}_{N} (\mathbf{\Theta}\circ\mathbf{U}) \mathbf{F}_{L}^T\mathbf{\Omega}_{\beta}^T\overline{\mathbf{F}}_{K},
\end{equation}
where $\circ$ denotes the Hadamard product, and $\overline{\mathbf{F}}$, $\mathbf{F}^T$ and $\mathbf{F}^*$ refer to the conjugate, transpose and conjugate transpose of the matrix ${\mathbf{F}}$, respectively. Let ${\mathbf{z}_2^{\prime}}=\mathrm{vec}({\mathbf{Z}_2^{\prime}})$ and ${\mathbf{u}}=\mathrm{vec}({\mathbf{U}})$ be the column-wise vectors of matrices ${\mathbf{Z}_2^{\prime}}$ and ${\mathbf{U}}$, respectively. The linear system (\ref{s3e2}) can be reformulated into a standard matrix-vector form
\begin{equation}\label{zmu}
   \mathbf{z}^{\prime}_2=\mathbf{Mu},
\end{equation}
where
\begin{equation}\label{m}
	\mathbf{M}=(\mathbf{F}_{K}^* \mathbf{\Omega}_{\beta}{\mathbf{F}}_{L}\otimes\mathbf{F}_{M}^*\mathbf{\Omega}_{\alpha}\mathbf{F}_{N} ) \mathrm{diag}(\mathbf{\Theta}),
\end{equation}
In (\ref{m}), $\otimes$ refers to the Kronecker product, and $\mathrm{diag}(\mathbf{\Theta})$ denotes the diagonal matrix that has the diagonal element $\mathrm{vec}(\mathbf{\Theta})$. It can be seen that the size of  $\mathbf{M}$  is  $MK\times NL$, and $MK<NL$ because the low-resolution image has a smaller size. Then (\ref{zmu}) is an underdetermined system. So, different from the conventional  interferometry problem, interferogram formation from (\ref{zmu}) is an ill-posed problem and thus cannot be solved directly.
\section{Interferogram Formation}

In virtue of compressive sensing theory, we now recover the interferogram by solving a sparse recovery problem.

\subsection{Interferogram Formation via Compressive Sensing}

 We use a wavelet basis $\bf W$ to sparsely represent the interferogram as $\mathbf{u}=\mathbf{W x}$, where $\bf x$ is a sparse coefficient vector.
Incorporating this representation into (\ref{zmu}) and defining a matrix
	$\mathbf{A}=\mathbf{M W}$,
we can describe the underdetermined system (\ref{zmu})  by a compressive sensing model
\begin{equation}\label{s3e3}
	{\mathbf{z}_2^{\prime}}=\bf A x,
\end{equation}
where $\bf A$ is called the sensing matrix.
According to compressive sensing theory, if the vector $\bf x$ is reasonably sparse and the sensing matrix $\bf A$ has small mutual coherence, we can recover $\bf x$ by solving the following optimization problem \cite{Donoho2006Compressed}
\begin{equation}\label{bpdn}
	\min\limits_{\mathbf{x}}^{}   \left\|\mathbf{z}_2^{\prime}-\mathbf{A}\mathbf{x}\right\|_2^2+\lambda\norm{\mathbf{x}}_1,
\end{equation}
where $\lambda$ is a regularization parameter used to tradeoff between the sparsity of $\mathbf{x}$ and the modeling accuracy. For the problem (\ref{bpdn}), there exist a variety of fast solvers, such as the \emph{fast iterative shrinkage-thresholding algorithm} (FISTA) \cite{beck2009fast}.

In the context of compressive sensing, the sensing matrix usually is required to have small mutual coherence, such that the recovery error can be well bounded \cite{candes2008introduction}.
In the following, we will analyze the sensing matrix $\bf A$ to show that it is expected to  have small mutual coherence.

\subsection{Speckle Effect and Random Sensing Matrix }

\begin{figure}
	\centering
	\subfigure{
		\includegraphics[width=1.3cm,height=4cm]{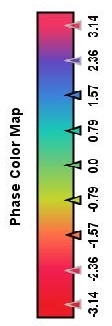}}
	\subfigure{
		\includegraphics[width=6cm,height=4cm]{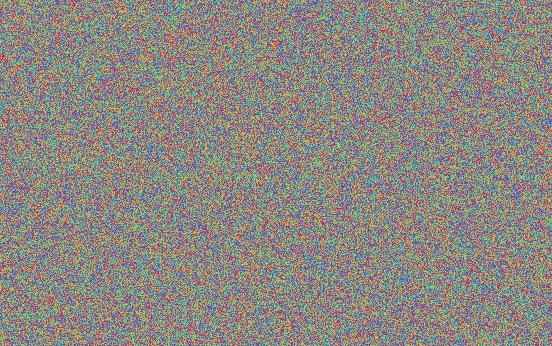}}\\
	\caption{Random phase of a SAR image (Copernicus Sentinel data [2018]). Due to the speckle effect, the scattering phase is randomly distributed in $(-\pi,\pi)$. The random phase leads to a random sensing matrix.
	}\label{ap}
\end{figure}

To get small mutual coherence, injecting randomness has taken the leading role in the design of sensing matrices \cite{Foucart2013A}.
For our model (\ref{s3e3}), the sensing matrix $\bf A$ contains natural randomness: due to the {speckle effect}, the phase of $\theta[n,l]$ is a independent random variable uniformly distributed in $(-\pi,\pi)$. Hence, the matrix $\mathbf{\Theta}$ can be viewed as a random matrix with i.i.d. entries uniformly distributed in the unit circle of the complex field. An example illustrating the random phase of a SAR image is shown in Fig. \ref{ap}. Then from  (\ref{s3e2}), we find that our model actually consists of random modulation by $\mathbf{\Theta}$,  filtering and resampling by 2-D DFT, lowpass filter and 2-D inverse DFT. This structure is similar to  \emph{random demodulation}  \cite{Kirolos2007Analog}, which is  an effective compressive sensing system. Therefore, the sensing matrix $\bf A$  is expected to  have small mutual coherence according to the compressive sensing theory.  The strict mathematical analysis of our model is in progress and we only validate it by numerical experiments in this paper.
It is  notable that the proposed model does not need artificially injecting randomness, which is comprehensively exploited in compressive sensing-based applications, and thus avoids complicating the radar data acquisition.

\section{Numerical Experiments}

We test the proposed method by several idealized numerical experiments using Sentinel-1 single look complex (SLC) images (online available at https://scihub.copernicus.eu/dhus). Results processed by our method using a high-resolution image and a low-resolution one, are compared with that processed by the conventional method  using two high-resolution images. The relative root mean squared error (RRMSE)  are used to evaluate the performance of our method.
The RRMSE is defined by the unwrapped phase $\phi_{rec}[n,l]$ and a reference phase $\phi_{ref}[n,l]$ as follows
\begin{equation}\label{}
  \mathrm{RRMSE}=10\,\mathrm{log}\left\{{\frac{\sum\limits_{n,l}{\left(\phi_{rec}[n,l]-\phi_{ref}[n,l]\right)^2}}{\sum\limits_{n,l}{\phi^2_{ref}[n,l]}}}\right\}.
\end{equation}

\subsection{Setup}
The used  Sentinel-1 SLC images contains $4096\times 4096$ pixels, which are collected on 2017-11-22 and 2017-12-04, respectively, and the illuminated area is located in  Iran, near Kavir desert. The images are collected in stripmap mode with $5\times 5\,$m spatial resolution. To generate a low-resolution image that is equivalent to the acquisition using a smaller bandwidth, the second image is filtered by an ideal lowpass filter, followed by resampling.

After forming two images with dual resolutions, we recover the interferogram by solving (\ref{bpdn}). We choose the Daubechies-4 wavelet basis to sparsely represent the  interferogram and set the regularization parameter to $0.0001$. Then we use FISTA as the recovery algorithm and set its maximum number of iterations to $200$.
Next, the interferogram returned by FISTA is filtered using  Goldstein filter and unwrapped by the SNAPHU software \cite{snaphu}. The filtering and unwrapping uses the default parameters in the SNAP software (online available at http://step.esa.int/main/snap-5-0-released/).
\subsection{Results and Discussions}
\begin{figure*}[!ht]
  \centering
\centering
\begin{overpic}[width=1.5cm,height=4.3cm]{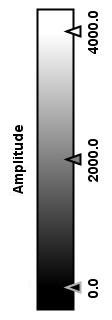}\end{overpic}
\begin{overpic}[width=4.3cm]{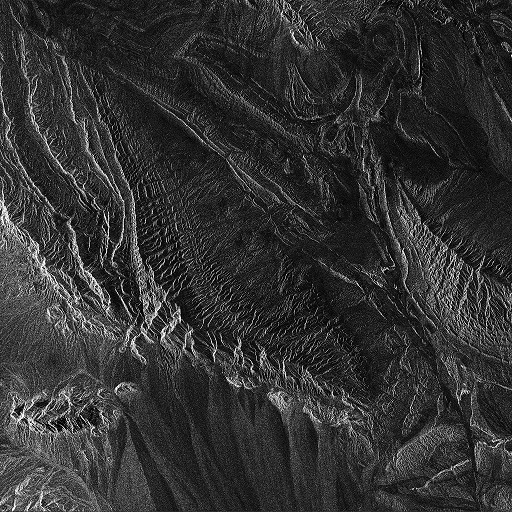}\end{overpic}\hspace{0.6em}
\begin{overpic}[width=4.3cm]{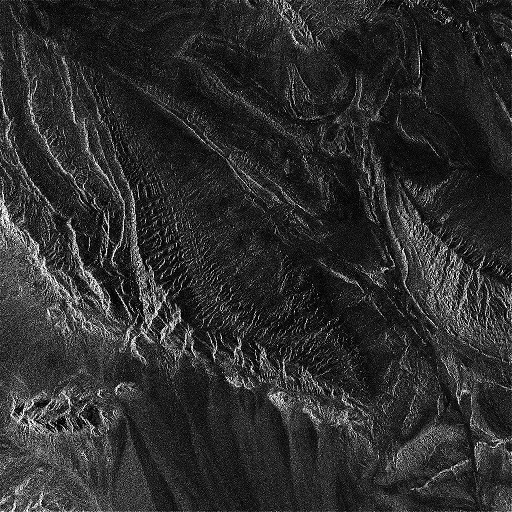}\end{overpic}\hspace{0.6em}
\begin{overpic}[width=4.3cm]{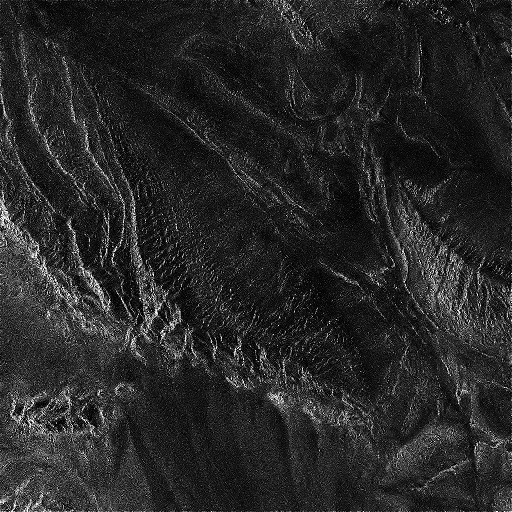}\end{overpic}\par\vspace{0.9em}
\begin{overpic}[width=1.5cm,height=4.3cm]{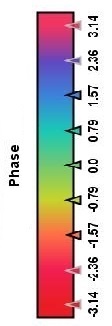}\end{overpic}
\begin{overpic}[width=4.3cm]{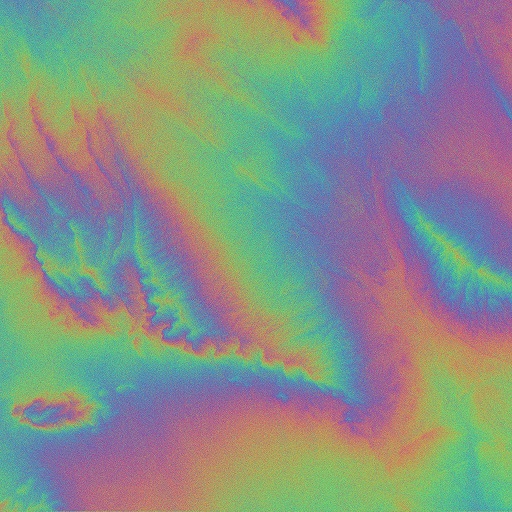}\end{overpic}\hspace{0.6em}
\begin{overpic}[width=4.3cm]{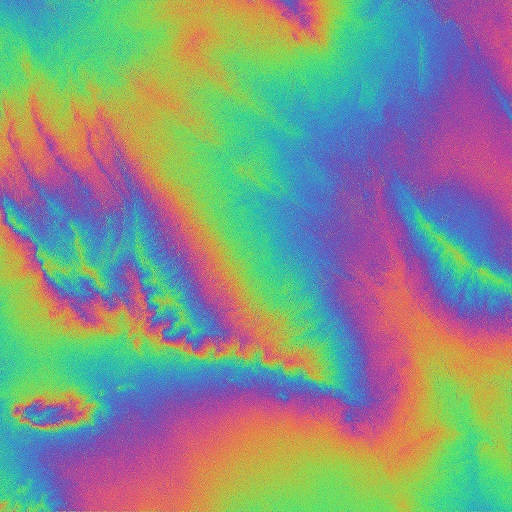}\end{overpic}\hspace{0.6em}
\begin{overpic}[width=4.3cm]{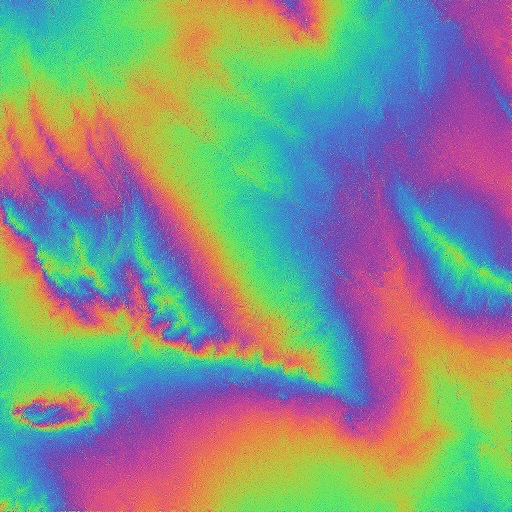}\end{overpic}\par\vspace{0.9em}
\begin{overpic}[width=1.5cm,height=4.3cm]{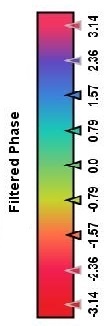}\end{overpic}
\begin{overpic}[width=4.3cm]{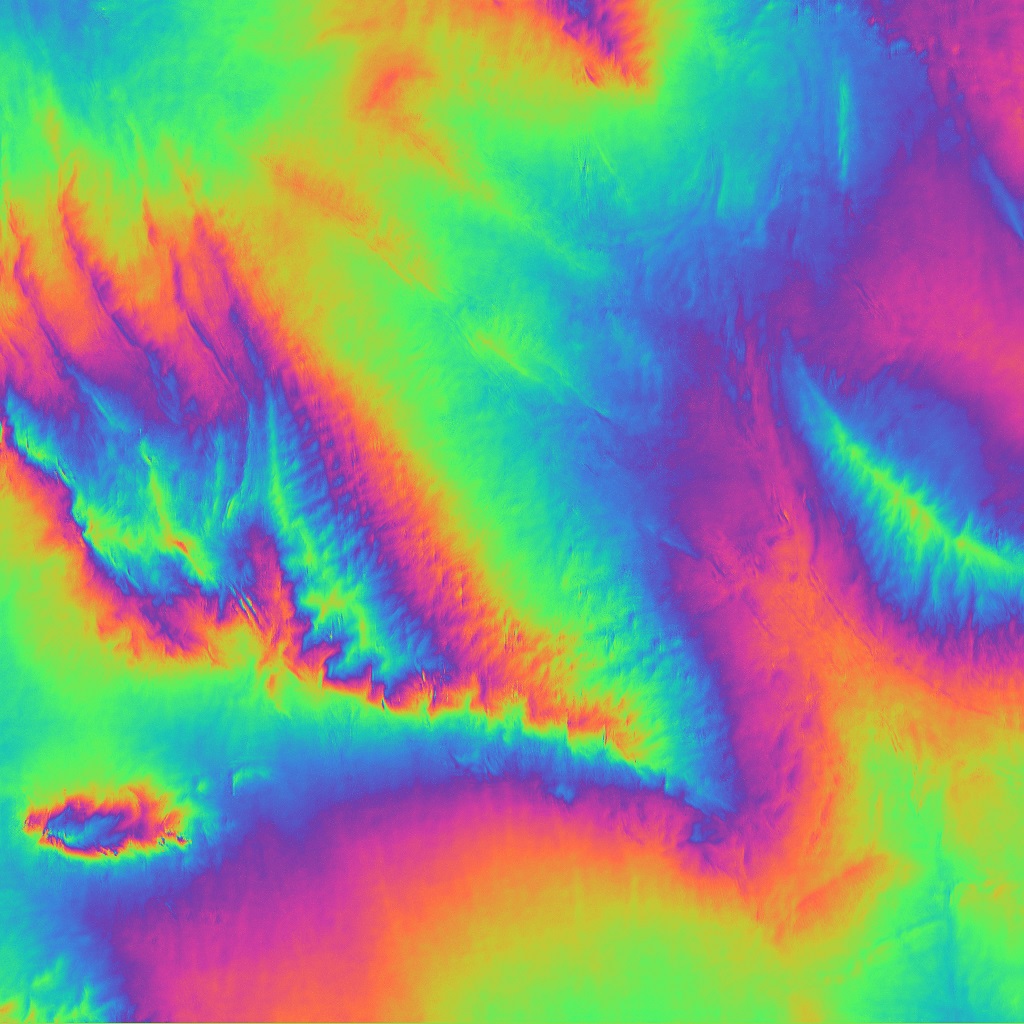}\end{overpic}\hspace{0.6em}
\begin{overpic}[width=4.3cm]{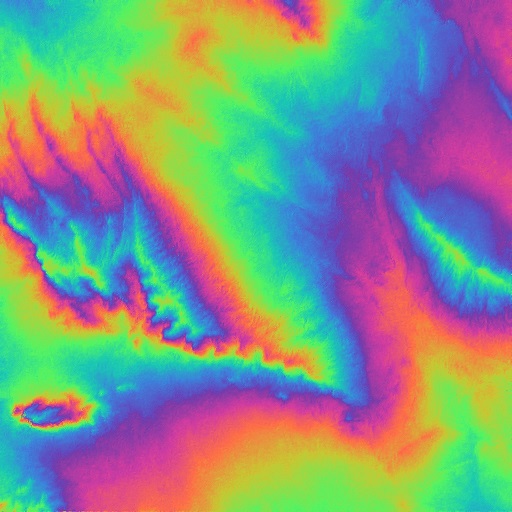}\end{overpic}\hspace{0.6em}
\begin{overpic}[width=4.3cm]{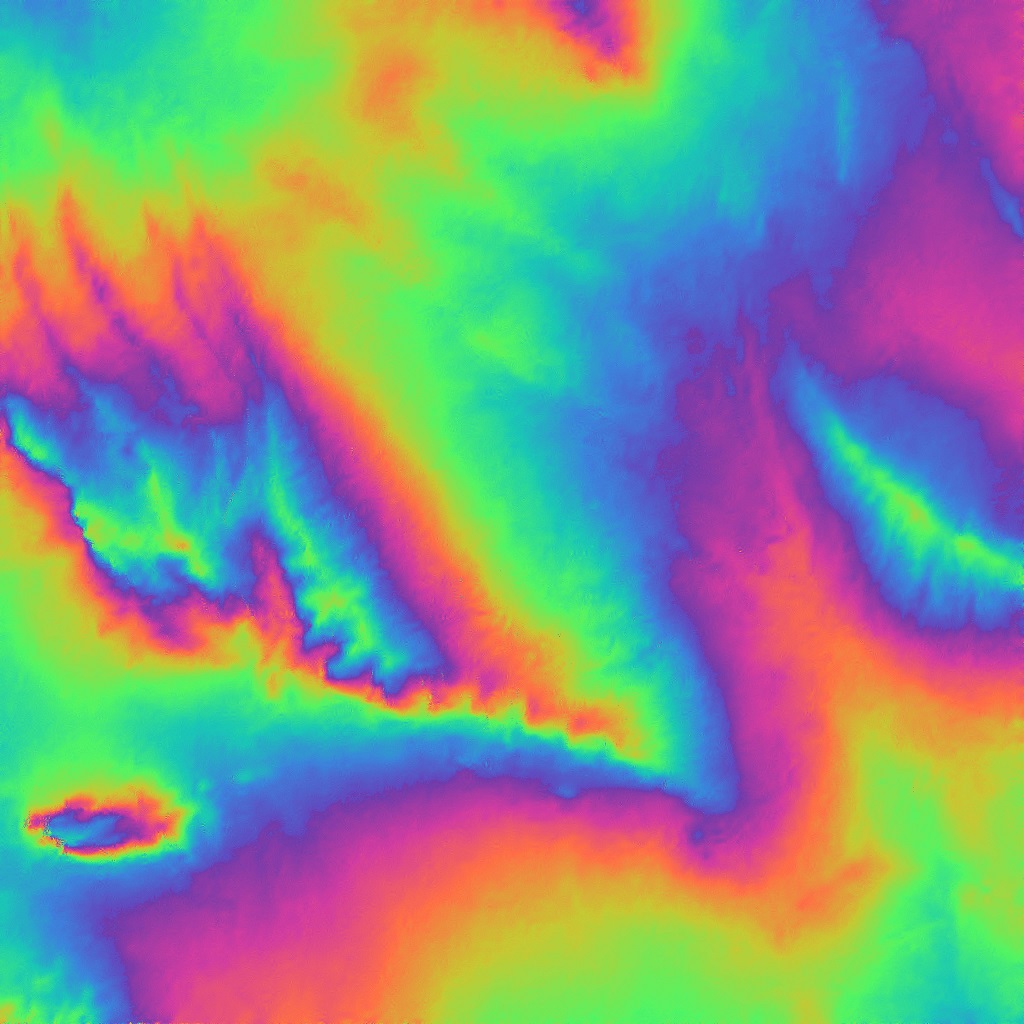}\end{overpic}\par\vspace{0.9em}
\hspace{0.35em}\begin{overpic}[width=1.5cm,height=4.3cm]{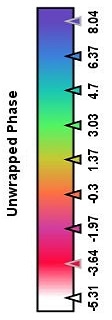}\end{overpic}
\begin{overpic}[width=4.3cm]{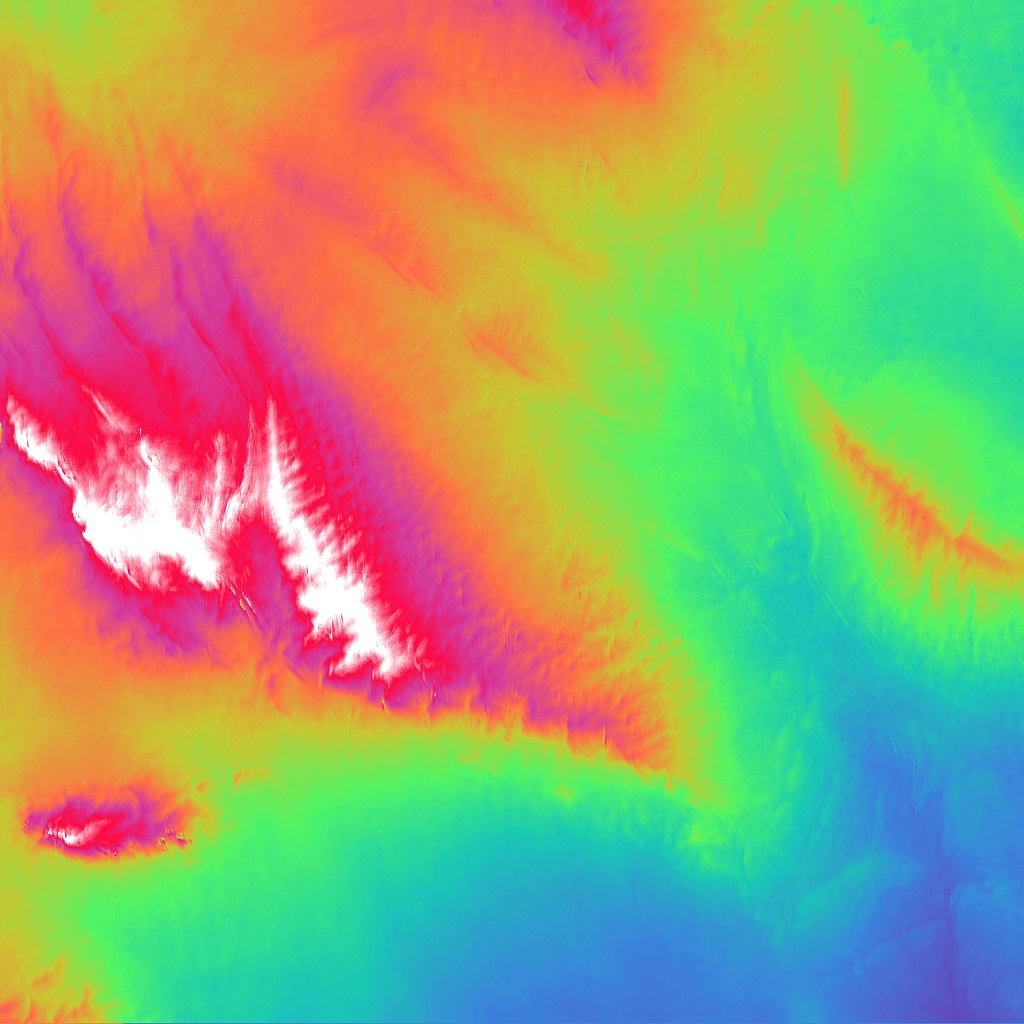}
			\put(40,-12){{\footnotesize$1\times 1$}}
		\end{overpic}\hspace{0.6em}
\begin{overpic}[width=4.3cm]{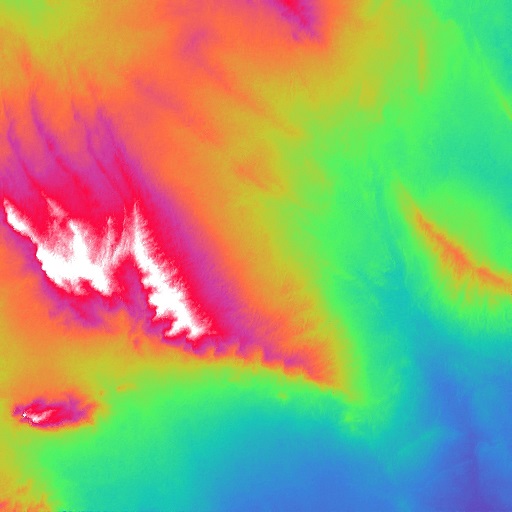}
			\put(35,-12){{\footnotesize$1/2\times 1/2$}}
		\end{overpic}\hspace{0.6em}
\begin{overpic}[width=4.3cm]{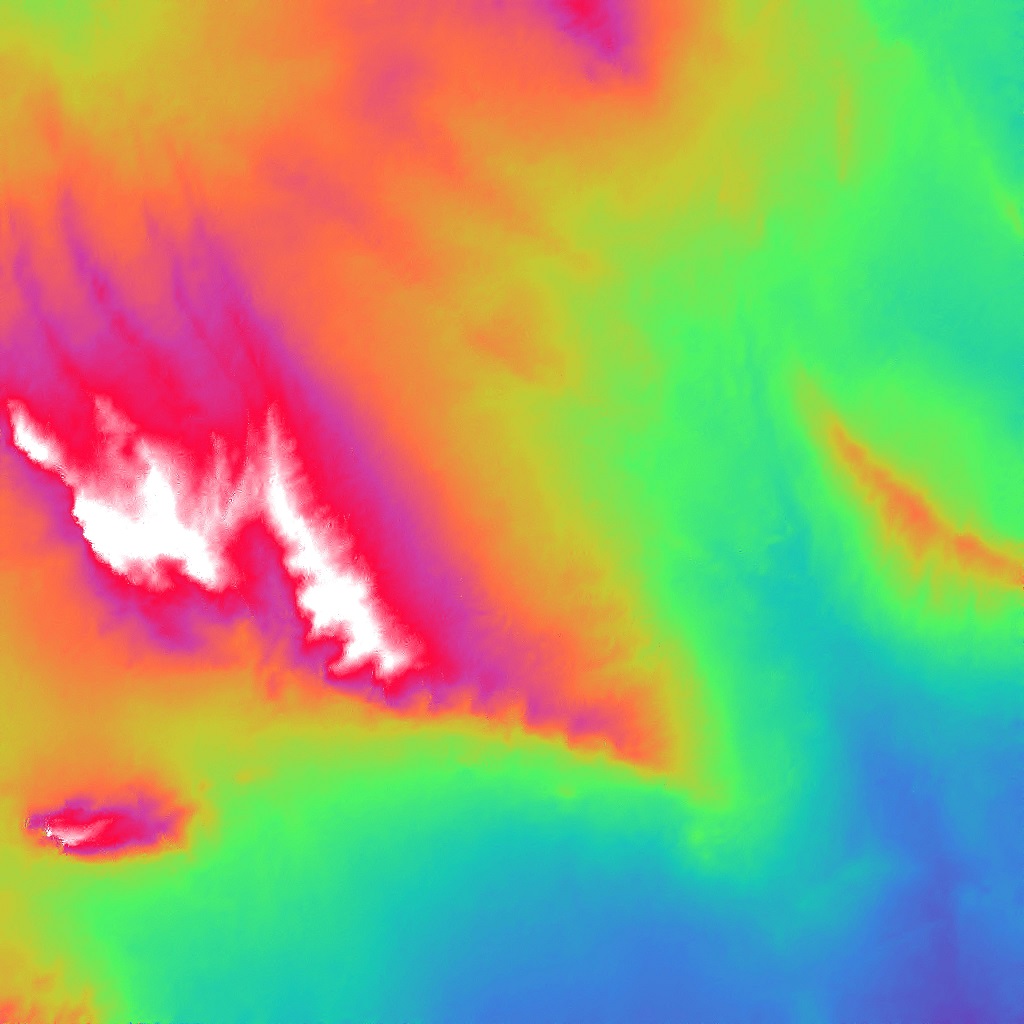}
			\put(35,-12){{\footnotesize$1/5\times 1/5$}}
		\end{overpic}
\vspace{1em}
	\caption{ Results processed by the conventional method using two high-resolution images (column 1), and those by our method using a high-resolution image and a low-resolution one (columns 2 and 3). The settings of $\alpha\times\beta$ are shown at the bottom of the figure.  Copernicus Sentinel data [2017]. }\label{phase}
\end{figure*}

Fig. \ref{phase} shows the interferogram formed by the conventional method using two high-resolution images and those by our method using a high-resolution image and a low-resolution one. As shown in the first and second rows of Fig. \ref{phase}, although the second interferometric image has low resolutions, our method can still recover the high-resolution interferogram with good visual qualities. Moreover, we find that our method results in cleaner interferograms, because  (\ref{bpdn}) naturally has the ability of denoising \cite{Elad2006Image}.

\begin{table}[!t]
	\renewcommand{\arraystretch}{1.3}
	\caption{Unwrapped phases}
	\label{tab1}
	\centering
	\begin{tabular}{c|c|c}
		\hline
		\bfseries ${\alpha}\times{\beta}$   &  $1/2 \times 1/2$ &  $1/5 \times 1/5$    \\
		\hline
		RRMSE                & -28.7dB    & -25.1dB   \\
		\hline
	\end{tabular}
\end{table}

Since the interferogram formed from the two high-resolution images is noisy, which usually needs filtering to reduce noise for phase unwrapping, direct comparison of these interferograms is not fair. As an alternative, we compare the filtered and  unwrapped phases.
As seen in the third and fourth rows of Fig. \ref{phase}, the filtered and  unwrapped phases of our results are almost the same as the conventional results.
To quantitatively measure our results, we take the unwrapped phase processed by the conventional method as the reference (the first image in row 4 of Fig. \ref{phase}), and calculate the RRMSE of our unwrapped phases (the second and third images in row 4 of Fig. \ref{phase}).  As shown in Table I, when we reduce the resolution of the second image to $1/5\times 1/5$ of its original values in range and azimuth dimensions, the resulting unwrapped phase still have good qualities, i.e., $-25.1$dB RRMSE. These numerical experiments demonstrate the effectiveness of our method.

\section{Conclusion and Future Works}
In this paper, we proposed a new method for effective extraction of interferometric information from radar images with different resolutions. We found that the underlying high-resolution interferogram is embedded into an underdetermined system. Then interferogram formation is transformed into a standard sparse recovery problem in compressive sensing by exploiting the sparsity of interferogram in wavelet domain. We found that the sensing matrix of our model is expected to  have small mutual coherence due to the speckle effect. Thus the recovery performance will be guaranteed according to the compressive sensing theory. Numerical experiments on Sentinel-1 data were made to illustrate the promising performance of our method.
However, we should point out that our method needs more computing resources to solve the compressive sensing-based interferometry problem.

Currently, theoretical analysis as well as some potential applications, including multi-mode SAR interferometry and deferential SAR interferometry, are under our research.

\bibliographystyle{IEEEtran}
\bibliography{IEEEabrv,mybib}

\end{document}